\documentclass[12pt]{article}

\usepackage{epsfig}
\usepackage{amsfonts}
\def\+{{+\!\!\!+}}

\def\d{\partial} 
 
\def\a{\alpha} 
\def\th{\theta}

\def\N{\nabla}

\def\r{\rho} 
\def\l{\lambda} 
 
\def\L{\Lambda}

\def\F{\Psi} 
\def\e{\varepsilon}

\def\pmb#1{\setbox0=\hbox{#1}%
\kern.0em\copy0\kern-\wd0 
\kern-.04em\copy0\kern-\wd0 
\kern.08em\copy0\kern-\wd0 
\kern-.04em\raise.0433em\box0 }         
 
\def\half{\frac{1}{2}}

\newcommand{\nc}{\newcommand} 
\nc{\beq}{\begin{equation}} 
\nc{\eeq}[1]{\label{#1}\end{equation}} 
\nc{\ber}{\begin{eqnarray}} 
\nc{\eer}[1]{\label{#1}\end{eqnarray}} 
\nc{\pek}[1]{\cite{#1}} 
\nc{\enr}[1]{(\ref{#1})} 
\nc{\kal}[1]{{\cal{#1}}} 
\nc{\dott}{\;\cdot\;} 

\def\0 {\nonumber}

\begin{document} 
\setcounter{page}{0}
\newcommand{\inv}[1]{{#1}^{-1}} 
\renewcommand{\theequation}{\thesection.\arabic{equation}} 
\newcommand{\be}{\begin{equation}} 
\newcommand{\ee}{\end{equation}} 
\newcommand{\bea}{\begin{eqnarray}} 
\newcommand{\eea}{\end{eqnarray}} 
\newcommand{\re}[1]{(\ref{#1})} 
\newcommand{\qv}{\quad ,} 
\newcommand{\qp}{\quad .} 
\begin{titlepage} 
\begin{center} 
                             
                             \hfill   hep-th/0409250\\ 
                             \hfill   UUITP-20/04\\
                             \hfill   HIP-2004-46/TH\\
                             
\vskip .3in \noindent 


{\large \bf{Generalized complex geometry and supersymmetric non-linear sigma models}} \\

\bf{Talk at Simons Workshop 2004}
\vskip .2in 

{\bf Ulf~Lindstr\"om$^{ab}$,}\\ 

\vskip .05in 



$^a${\em\small Department of Theoretical Physics\\ 
Uppsala University, Box 803, SE-751 08 Uppsala, Sweden} \\
\vskip .02in
$^b${\em\small HIP-Helsinki Institute of Physics\\
P.O. Box 64 FIN-00014 University of Helsinki, Suomi-Finland}\\


\vskip .5in 

\vskip .1in 
\end{center} 

\begin{center} {\bf ABSTRACT }  
\end{center} 
\begin{quotation}\noindent  
After an elementary presentation of the relation between supersymmetric 
nonlinear sigma models and geometry, I focus on $2D$ and the target space geometry
allowed when there is an extra supersymmetry. This leads to a brief introduction to
generalized complex geometry, a notion introduced recently by Hitchin which
interpolates between complex and symplectic manifolds.  Finally I  present worldsheet 
realizations of this geometry,

\end{quotation} 
\vfill 
\eject 


\end{titlepage}

\section{Introduction}

A year ago at the previous Simons workshop Marco Gualtierei presented
part of what later became his thesis entitled ``Generalized Complex Geometry'' (GCG)
\cite{Gualtieri}. It was based in work done by his supervisor, Nigel Hitchin , who had 
been mainly motivated by describing generalized Calabi Yau manifolds, e.g., when including 
an antisymmetric $B$-field \cite{Hitchin}. Another interesting aspect, however, is the relation 
to supersymmetric non-linear sigma models. The geometry he described includes the bi-hermitean
geometry with a $B$-field found by Gates, Hull and Ro\v cek 20 years ago \cite{Gates:nk}.
Indeed, it is an interesting fact that still 25 years after the original classification by 
Zumino \cite{Zumino:1979et} 
there are still some open problems in this area. Both the hyperk\"ahler geometry discussed by 
Alvarez-Gaum\'e and Freedman \cite{Alvarez-Gaume:1980vs},
and the bi-hermitean geometry of \cite{Gates:nk} correspond to additional supersymmetries that in general 
only close on-shell. This prevents formulation with all the supersymmetries manifest, except in special cases.
In addition, although the $B$-field does have a geometrical role in the $\kal{N}=(2,2)$ 
sigma model as a potential for the torsion, it is only locally defined  and the model really only depends on its 
field strength.  It would be nice to have a geometrical setting where the $B$-field itself aquires a geometrical meaning.

It seemed that the GCG of Hitchin may help shed light on some of these questions, and this is a report on some subsequent
development in that direction partly in collaboration with R. Minasian, A.Tomasiello 
and M.Zabzine \cite{Lindstrom:2004eh}, \cite{Lindstrom:2004iw}. Originally my talk was to 
be the first of two on this subject, the second was to be delivered by Maxim Zabzine. 
Unfortunately, due to the vagaries of the
US consular system, he was unable to attend the workshop.

\section{Sigma models}

A non-linear sigma model is a theory of maps
\ber
&&X^\mu(\xi):\kal{M}\rightarrow \kal{T}~,
\eer{map}
where $\xi^{i}$ are coordinates on $\kal{M}$ and $X^\mu(\xi)$ coordinates on the target space 
$\kal{T}$. Classical solutions are found by extremizing the action.
\ber
&&S=\int d\xi ~\partial_{i}X^\mu G_{\mu\nu}(X)\partial^{i}X^\mu~,
\eer{act}
where the symmetric $G_{\mu\nu}$ is identified with a metric on $\kal{T}$, a first sign 
of the intimate relation between sigma models and target space geometry. Extremizing $S$
results in the $X^\mu$'s being harmonic maps involving the pull-back of the covariant Laplacian:
\ber
&&\N^{i}\partial_{i}X^\mu=0~,
\eer{har}
where $\N$ is defined w.r.t. the Levi-Civita connection for $G$, another indication of the relation to geometry.
The geometry is Riemannian for the bosonic model, but typically becomes complex when we impose 
supersymmetry. 

Supersymmetry is introduced by replacing the $X^\mu$'s by superfields:
\ber
&&X^\mu(\xi)\to\phi^\mu(\xi,\th)~,
\eer{suf}
with component expansion
\ber
&&X^\mu=\phi^\mu |\cr
&&\F^\mu_{\a}=D_{\a}\phi^\mu |\cr
&&.........~,
\eer{com}
where $D_{\a}$ are the superspace spinorial covariant derivatives generating the supersymmetry algebra, $|$
denotes ``the $\th$ indepentent part of«« and ``....««indicates additional components depending on the dimension 
of $\kal{M}$. To be more concrete, in $4D$, using Weyl spinors, the supersymmetry algebra is
\ber
&&\{D_{\a},\bar D_{\dot\a}\}=2i\partial_{\a\dot\a}~,
\eer{alg}
and the smallest representation containing a scalar field is a chiral superfield $\bar D_{\dot\a}\phi
=D_{\a}\bar\phi=0$. This means that the target space will naturally have complex coordinates.
The most general supersymmetric action is determined by an arbitrary function $K$
\ber
&&S=\int d^4\xi d^2\th d^2\bar \th K(\phi,\bar\phi)=\int d^4\xi\left(
\frac {\partial^2K}{\partial \phi^\mu\partial\bar\phi^\nu}
\partial^{i}X^\mu\partial_{i}
\bar X^\nu+....\right)¥~,
\eer{sact}
and a comparison to (\cite{act}) shows that $K$ is a K\"ahler potential. The geometry is thus
K\"ahler with metric $g_{\mu\bar\nu}=\partial_{\mu}\partial_{\bar\nu}K$. 

A similar analysis in other dimensions leads to a classification for dimensions \footnote{Higher $D$'s
will necessarily have multiplets with vector components.}$D\leq 6$ which may be summarized as 
\footnote{For brevity only even $D$'s are included.}
\bigskip
\begin{center}¥
\begin{tabular}[htb]{|l|lll|l|}
\hline
D&6&4&2&GEOMETRY\cr
\hline
N&1&2&4&Hyperk\"ahler\cr
&&1&2&K\"ahler\cr
&&&1&Riemannian\cr
\hline
\end{tabular}
\end{center}¥
\bigskip

The interesting part, where this table is ``incomplete'' is for $D=2$, the dimension relevant for string theory.
There are (at least) two special features in $D=2$. First, there can be different amounts of supersymmetry in the 
left and right moving sectors denoted $\kal{N}=(p,q)$ supersymmetry. Second, if parity breaking terms are allowed,
the background may contain an antisymmetric $B_{\mu\nu}$-field.
For $\kal{N}=(2,2)$, the supersymmetric action written in terms of real $\kal{N}=(1,1)$ superfields  reads
\ber
S=\int d^2\xi d^2\th d^2 D_{+}\phi^\mu E_{\mu\nu}(\phi)D_{-}\phi^\nu~,
\eer{sac2}
where $E_{\mu\nu}(\phi)\equiv G_{\mu\nu}(\phi)+B_{\mu\nu}(\phi)$. This action has manifest 
$\kal{N}=(1,1)$ supersymmetry without any additional restrictions on the target space geometry. 
Gates, Hull and Ro\v cek showed that it has an additional non-manifest supersymmetry,
\ber
&&\delta\phi^\mu= \e^+D_{+}\phi^\nu J_{\mu}^{(+)\nu}+\e^-D_{-}\phi^\nu J_{\mu}^{(-)\nu}~,
\eer{susy2}
provided that the following conditions are fulfiled\footnote{There is a further relation between the torsion 
and the complex structures. which we left out}.¥:\\
$\bullet $ Both the $J$'s are {\em almost complex structures}, i.e. $J^{(\pm)2}=-1$.\\
$\bullet $ They are {\em integrable}, i.e., their Nijenhuis-tensors vanish\\
\beq
\kal{N}_{\mu\nu}^{(\pm)~\r}\equiv  J_{\l}^{(\pm)\mu}\d_{[\l} 
J_{\nu]}^{(\pm)\r}-(\mu\leftrightarrow\nu) =0
\eeq{nij}
$\bullet $  The metric is {\em hermitean} w.r.t. both complex structures, i.e. they both preserve the metric
$\qquad J^{(\pm)t}GJ^{(\pm)}=G$\\
 $\bullet $ The $J$'s are {\em covariantly constant} with respect to a torsionful connection:\\
 $\qquad\N^{(\pm)}J^{(\pm)}=0$ with $\N^{(\pm)}\equiv \N ^0\pm H$, the Levi-Civita connection 
plus completely antisymmetric torsion in form of the field-strength $H=dB$\footnote{Strictly, the torsion $T=g^{-1}H¥$.}.

The above conditions represent a bi-hermitean target space geometry with a $B¥$-field, and result from 
requiring invariance of the action (\ref{sac2}) under the transformations (\ref{susy2}) as well as closure of 
the algebra of these transformations. Closure is only achieved on-shell, however. Only under the special 
condition that the two complex structures commute does the algebra close off-shell. In that case there is a
manifestly $\kal{N}=(2,2)$
action for the model, given in terms of chiral and twisted chiral $\kal{N}=(2,2)$ superfields \cite{Gates:nk}.
An interesting question is thus: What is the most general $\kal{N}=(2,2)$ sigma model (with off-shell closure of the algebra)
and what is the corresponding geometry? In asking this we  have in mind an extension of the model to include 
additional fields to allow off-shell closure in the usual ``auxiliary field'' pattern and a geometry that includes 
these fields.

As mentioned in the introduction, the GCG does contain the bi-hermitean geometry as a special case and thus seems
a promising candidate. We therefore turn to a brief description of the GCG.

\section{Generalized Complex Geometry}

To  understand the generalization, let us first briefly look at some aspects of the definition of the ordinary complex 
structure. The features we need are that an almost complex structure $J$ on a $d$-dimensional 
manifold $\kal{T}¥$¥ is a map from the tangent bundle
$J: T\to T$ that squares to minus the identity $J^2=-1$. With these properties $\pi_{\pm}\equiv \frac 1 2 (1\pm iJ)$
are projection operators, and we may ask when they define integrable distributions. The condition for this is that
\beq
\pi_{\mp}[\pi_{\pm}X,\pi_{\pm}Y]=0
\eeq{int}
for $X,Y\in T$ and $[,¥]¥$ the usual Lie-bracket on $T$. This relation is equivalent to the vanishing of 
the Nijenhuis tensor $\kal{N}(J)$, as defined in (\ref{nij}).

To define GCG, we turn our attention from the tangent bundle $T(\kal{T}¥)¥$ to the 
sum of the tangent bundle and the co-tangent 
bundle $T\oplus T^*$. (Note that the structure group of this bundle is $SO(d,d)$, the string theory
T-duality group, an important fact that will not be further pursued in this lecture).
We write an element of $T\oplus T^*$ as $X+\xi$ with the vector $X\in T$ and the one-form
$\xi\in T^*$. The natural pairing $(X+\xi ,X+\xi)=-\imath _{X}\xi$ gives a metric \kal{I} on 
$T\oplus T^*$ as $X+\xi$, which in a coordinate basis $(\d_{\mu} ,dx^\nu)$ reads
\beq
\left(\begin{array}{cc}
0&1_{d}\cr
1_{d}&0\end{array}\right)~.
\eeq{Imet}
In the definition of a complex structure above we made use of the Lie -bracket on $T$. To define 
a generalised complex structure we will need a bracket on $T\oplus T^*$. The relevant bracket is the 
the skew-symmetric {\em Courant bracket} \cite{courant} defined by
\beq
[X+\xi , Y+\eta]_{c}\equiv [X,Y]+\pounds_{X}\eta-\pounds_{Y}\xi
-\half d(\imath _{X}\eta -\imath _{Y}\xi)~.
\eeq{cour}
This bracket equals the Lie-bracket on $T$ and vanishes on $T^*$. \footnote{It 
does not in general satisfy the Jacobi 
identity; had it satisfied the Jacobi 
identity $(T\oplus T^*, [,]_{c})$ would have formed a Lie algebroid. It {\em does} satisfy
the Jacobi identity on subbundles  $L\subset T\oplus T^* $ that are Courant involutive and
 isotropic w.r.t. $\kal{I}$, but fails to do so in general. It fails in an interesting way which leads to the definition of 
 a Courant algebroid \cite{Gualtieri}.}  The most important property for us in the context of 
 sigma-models is that its group of automorphisms is not only $Diff(\kal{T})$ but also
 {\em b-transforms} defined by closed two-forms $b$,
 \beq
 e^b(X+\xi)\equiv X+\xi+\imath _{X}b~,
 \eeq{btf}
 namely,
 \beq
 [e^b(X+\xi), e^b(Y+\eta)]_{c}= e^b[X+\xi ,Y+\eta]_{c}~.
 \eeq{btf2}
 
 A {\em generalized almost complex structure} is an endomorphism 
 $\kal{J}: T\oplus T^*\to T\oplus T^*$ that satisfies 
 $\kal{J}^2=-1_{2d}$ and preserves  the natural metric $\kal{I}$,
 $\kal{J}^t\kal{I}\kal{J}=\kal{I}$. The projection operators $\Pi_{\pm}\equiv \frac 1 2 (1\pm i\kal{J})$ 
 are then used to define integrability (making $\kal{J}$¥ a generalized complex structure) as
 \beq
\Pi_{\mp}[\Pi_{\pm}(X+\xi),\Pi_{\pm}(Y+\eta)¥]_{c}¥=0
\eeq{int2}

In a coordinate basis $\kal{J}$ is representable as
\beq
\kal{J}=\left(\begin{array}{cc}
J&P\cr
L&K\end{array}\right)~,
\eeq{Jcor}
where $J:T\to T,\quad P:T^*\to T,\quad L:T\to T^*,\quad K:T^*\to T^*$. The condition 
$\kal{J}^2=-1_{2d}$ will impose conditions 
\ber
&&J^2+PL=-1_{d}¥\cr
&&JP+PK=0\cr
&&KL+LJ=0\cr
&&LP+K^2=-1_{d}~,
\eer{cond3}
and (\ref{int2}¥) will impose differential conditions on $J,P,L$ and $K$.

The ordinary complex structure is given by
\beq
\kal{J}_{J}¥=\left(\begin{array}{cc}
J&0\cr
0&-J^t\end{array}\right)~,
\eeq{JJ}
and a symplectic structure $\omega$ corresponds to
\footnote{For a generalized complex structure to exist $T$ has to be even-dimensional.}
\beq
\kal{J}_{\omega}¥=\left(\begin{array}{cc}
0&-\omega^{-1}\cr
\omega&0\end{array}\right)~.
\eeq{Jo}
A b-transform acts as follows
\beq
\kal{J}_{b}¥=\left(\begin{array}{cc}
1&0\cr
b&1\end{array}\right)\kal{J}\left(\begin{array}{cc}
1&0\cr
-b&1\end{array}\right).
\eeq{Jcor}

The general situation is illustrated in the following diagram:
\begin{center}
\epsfig{file=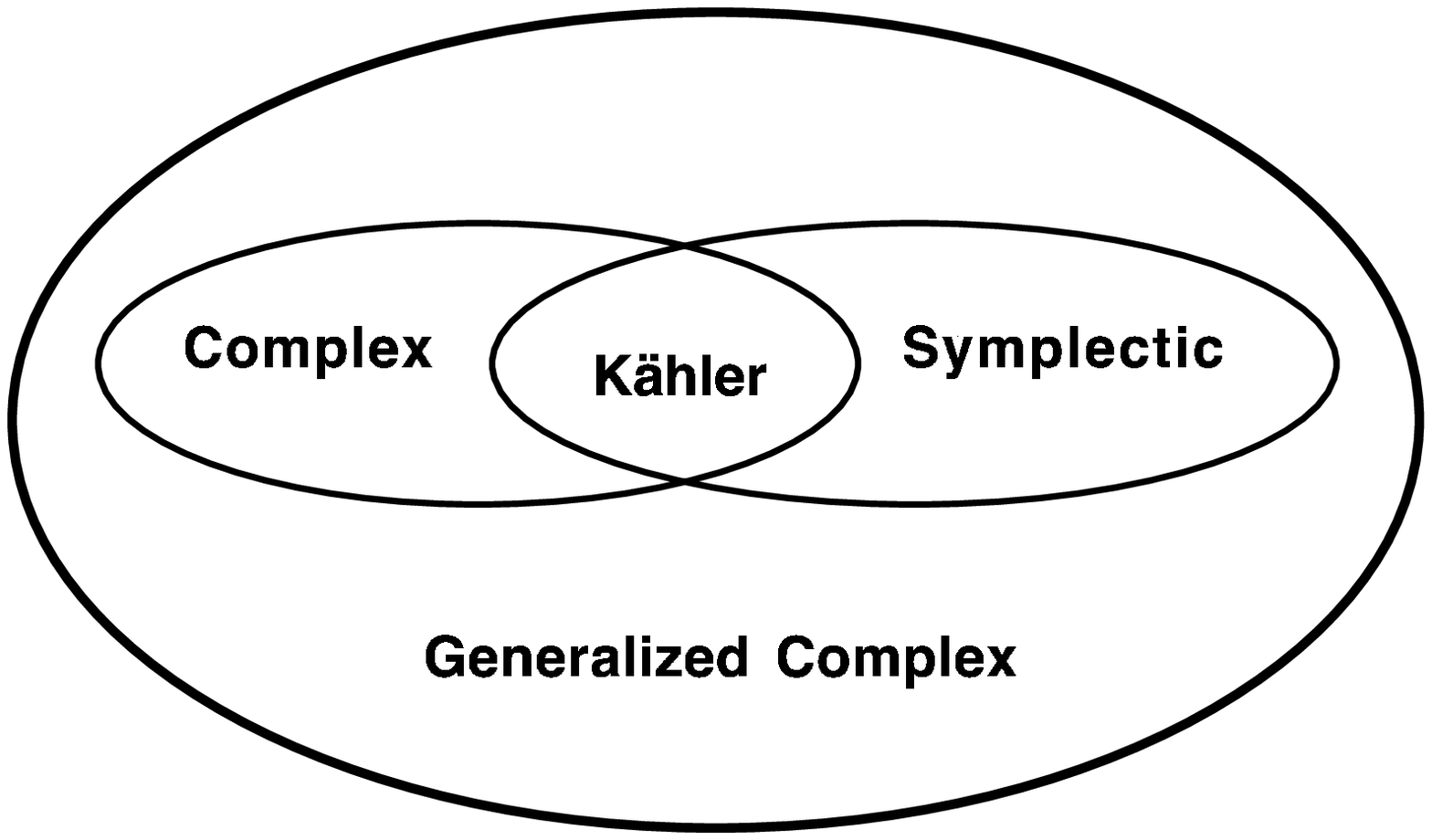, angle=0, width=8cm} 

Fig1. The relation between the different geometries discussed.
\end{center}¥
A useful property for calculations is that locally (in an open set around a non-degenerate point) 
 a manifold which 
admits a generalized complex structure may be brought to look like an open set in $\mathbb{C}^k$ 
times an open set in $(R^{2d-2k},\omega)$, where $\omega $ is in Darboux coordinates and $\mathbb{C}^k$
in complex (holomorphic and anti-holomorphic) coordinates (using diffeomorphisms and b-transform)
\footnote{The proof of this, generalizing the Newlander-Nirenberg and the Darboux theorems, may be found in 
Gualtieri's thesis, \cite{Gualtieri}, Sec. 4.7.}.
\begin{center}¥
\epsfig{file=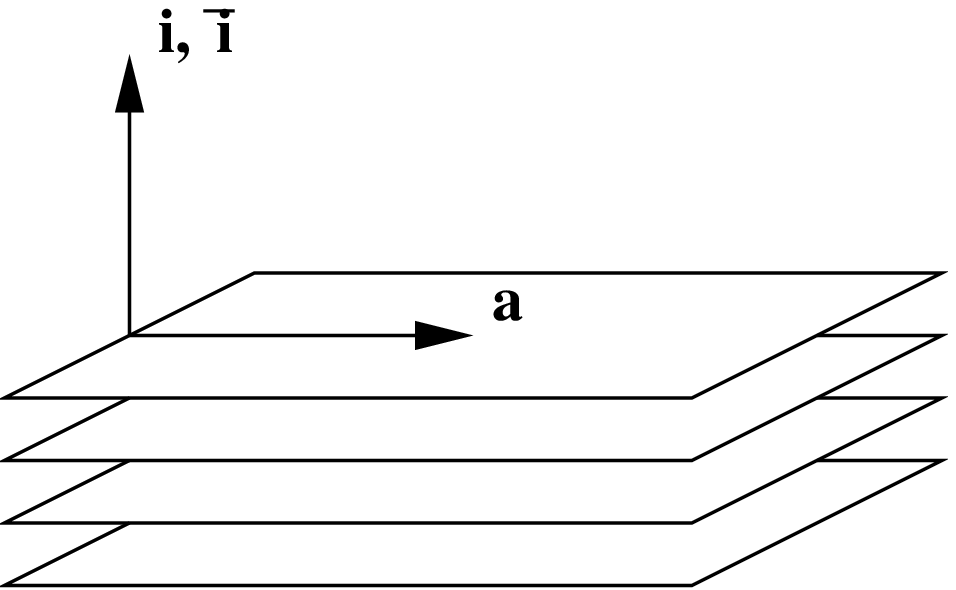, angle=0, width=8cm}
\vspace{1cm}¥

Fig 2. A naive picture  of the local foliation in complex coordinates $z^{i}, \bar z^{i}$\\
 and Darboux coordinates
$x^{a}¥$.
\end{center}

The generalized complex geometry is said to be generalized K\"ahler if there exist two commuting generalized 
complex structures $\kal{J}_{1}¥$ and $\kal{J}_{2}¥$ such that $G=-\kal{J}_{1}¥\kal{J}_{2}¥$ is a positive definite 
metric on $T\oplus T^*$. For a  K\"ahler manifold $(J,g,\omega)$, using (\ref{JJ}¥) and (\ref{Jo}¥) one finds
the metric
 \beq
G=-\kal{J}_{J}\kal{J}_{\omega}=\left(\begin{array}{cc}
0&g^{-1}\cr
g&0\end{array}\right)~.
\eeq{Jcor}

Finally, it is worth mentioning that it is possible to twist the above structure by a closed three-form.

We now turn to the question of how this geometry may be realized in sigma models.

\section{Sigma model realization}

In the sigma model action (\ref{sac2}) $D\phi\in T(\kal{T}¥)¥$.  Clearly we shall need a formulation with additional fields 
$S\in T^*$ to be able to realize the GCG. We thus consider the following first order action
\beq
S=\int d^2\xi d^2\th\left(S_{\mu +}E^{\mu\nu}(\phi)S_{\nu -}-S_{\mu (+}D_{-)}\phi^\mu +
D_{+}\phi^\mu (B-b)_{\mu\nu}¥D_{-}\phi^\nu\right)~,
\eeq{suac1}
where $E_{\mu\nu}\equiv g_{\mu\nu}+ b_{\mu\nu}$ and its inverse may be thought of as  open string data:
\ber
E^{(\mu\nu)}=G^{\mu\nu}~,\quad E^{[\mu\nu]}=\th^{\mu\nu}~.
\eer{opd}
In ({\ref{suac1}) $S_{\mu\pm}$ acts as an auxiliary field which extends the model to a sigma model on 
$T\oplus T^*$ and $b$ is a globally defined two-form which allow us to display the b-transform. 
(Note that the original model ({\ref{suac1}) depends only on $H=dB$, and $B$ is thus typically only locally defined).
Eliminating  $S_{\mu\pm}$ we recover the action in ({\ref{sac2}). The b-transform is the statement that 
if in two actions of the form ({\ref{suac1}) $E_{\mu\nu}$ and $\tilde E_{\mu\nu}$ differ by a closed two-form 
$\tilde b$ the two actions are equivalent. 

The action ({\ref{suac1}) has many interesting limits. For example, if the metric is set to zero it is a supersymmetric version of a 
Poisson sigma model \cite{Schaller:1994es}. In what follows we shall not be interesting in the difference between the $B$
and $b$-fields but set them equal each other, so the $\kal{N}=(1,1)$ action we study is
\beq
S=\int d^2\xi d^2\th\left(S_{\mu +}E^{\mu\nu}(\phi)S_{\nu -}-S_{\mu (+}D_{-)}\phi^\mu \right)~,
\eeq{suac2}
We shall also be interested in it's \kal{N}$=(1,0)$ reduction
\beq
S=\int d^2\xi d^2\th\left(S_{\mu +}E^{\mu\nu}(\phi)S_{\nu =}-S_{\mu +}\d_{=}\phi^\mu +
D_{+}\phi^\mu S_{\mu =}\right)~,
\eeq{suac2}
and the purely topological $\kal{N}=(1,0)$ model
\beq
S=\int d^2\xi d^2\th\left(S_{\mu +}\d_{=}\phi^\mu \right)~.
\eeq{top}
For the $\kal{N}=(1,1)$ model, the form of ansats for the second 
supersymmetry ($\delta =\delta^{(+)}+\delta^{(-)}$) is determined by a dimensional analysis to be 
\ber
&&\delta^{(\pm)}\phi^\mu=\e^{\pm}\L^{A}_{\pm}A^{(\pm)\mu}_{A}¥\cr
&&\delta^{(\pm)}S_{\mu \pm}=\e^{\pm}\left(D_{\pm}¥\L^{A}_{\pm}B^{(\pm)}_{\mu A}
+\L^{A}_{\pm}\L^{B}_{\pm}C^{(\pm)}_{\mu AB}\right)\cr
&&\delta^{(\pm)}S_{\mu \mp}=\e^{\pm}\left(D_{\pm}¥\L^{A}_{\mp}M^{(\pm)}_{\mu A}
+D_{\mp}\L^{A}_{\pm}N^{(\pm)}_{\mu A}
+\L^{A}_{\pm}\L^{B}_{\mp}X^{(\pm)}_{\mu AB}\right)~,
\eer{gsusy2}¥
where $L^{A}_{\pm}\equiv (D_{\pm}\phi^\mu ,S_{\mu \pm})$ and all the coefficient are fuctions of $\phi$.
The conditions which follow from invariance of the action and closure of the algebra are of two kinds, 
algebraic and differential. The two (A-type) index coefficients typically turn out to be given as derivatives
of the one-index coefficients, just like the generalized complex structures contain are given in terms of $J,P,L$ and
$K$ which subsequently obey differential conditions via the integrability requirement.

For the topological model, considering only the left-moving sector, the relations corresponding to (\ref{gsusy2}) 
simplify considerably. They are
\ber
&&\delta^{(+)}\phi^\mu=\e^+\left(D_{+}\phi^\l J^\mu_{~\l}-S_{\l +}P^{\mu\nu}\right)\cr
&&\delta^{(+)}S_{\mu +}=\e^{+}\left(i\d_{\+}\phi^\l L_{\mu\l}-D_{+}S_{\l +}K^{~\l}_{\mu}+...\right)~,
\eer{gsusy3}
where ``....'' indicates the higher coefficients determined by the differential conditions.
In \cite{Lindstrom:2004iw} we show that the algebraic and differential conditions in this case are satisfied if and only if
\bea \nonumber
\kal{J}=\left(\begin{array}{cc}
J&P\cr
L&K\end{array}\right)~,
\eea
is a generalized complex structure.

Similar results hold for the full $\kal{N}=(2,0)$ sigma model, but there we were not yet
able to find the most general solution to the differential constraints. Under certain assumptions, 
however, we found a solution which is the geometry given by the following generalized complex 
structure
\bea \nonumber
\kal{J}=\left(\begin{array}{cc}
J&0\cr
L&-J^t\end{array}\right)~,
\eea
with $L_{\mu\nu}=J^\r_{[\nu}b_{\mu ]\r}$ and $\N^{(+)}_{\mu}J_{\nu\r}=0$. 
All other components are again determined by $\kal{J}$.

The full $\kal{N}=(1,1)$ model presents the most challenge as the solutions to the conditions corresponding to 
($\kal{N}=(1,0)$) are least known. In \cite{Lindstrom:2004eh}, where the model was introduced, 
the relation to the bi-hermitean geometry was 
established.  However, some of the assumptions in that paper may be considerably relaxed, and there are reasons to
expect that this relaxation will be enough to make the model invariant with the algebra closing off-shell.
\eject

\section{Conclusions}

We have presented  GCG and shown how it can be realized in the context of supersymmetric nonlinear sigma models.
Since we have not yet found the general solution to the constraints on the transformation coefficients for $\kal{N}=(2,0)$
and $\kal{N}=(2,2)$, we cannot yet say that  GCG is the most general target space geometry. In fact we have found 
hints that the full solution of the constraints may go beyond GCG, but this is yet unclear.
Off-shell closure is not yet proven, but seems possible. That would certainly make the whole approach more interesting, 
since in that case not only would we have an nice geometrical framework, which reduces to the known
bi-hermitean geometry when the auxiliary $S$ is integrated out, but also the possibility of extending the action while
keeping  $\kal{N}=(2,2)$ supersymmetry. This would allow, e.g., inclusion of higher derivative terms, in keeping
with the sigma model as an effective action.  There are many other directions one could investigate starting from the new 
form of the action. For example, it allows the study of models where $E^{\mu\nu}$ is non-invertible. 
The question for open sigma models 
of what are the most general boundary conditions allowed by supersymmetry has proven very useful in understanding 
geometrical restrictions on $D$-branes \cite{Lindstrom:2002vp, Lindstrom:2002jb} and has a natural extension in to the present case.
In fact GCG has already helped in interpreting some of the geometric structures previously found \cite{Zabzine:2004dp}.

The question of which criteria would allow the $\kal{N}=(2,2)¥$ sigma models with a $B$-field to be used for constructing topological 
strings was raised during the talk. In analogy to the Calabi-Yau case, a natural conjecture is 
that the criterion is the vanishing of the first Chern-class, 
as defined by the torsionful Ricci tensors $R^{{\pm}¥}_{\mu\nu}¥$. Since the torsion enters the Ricci tensor quadratically, $R^{+}=R^{-}$, but there
are nevertheless two possible Ricci-forms depending on $J^{+}$ and $J^{-}$ respectively. 
Again, perhaps this question is best addressed within the framework
of generalized complex geometry. A recent relevant paper is \cite{kapustin}.

Finally, we mention that GCG has been considered in other contexts. Studying generalized Calabi Yau manifolds 
(the original motivation) and discussing supersymmetrical backgrounds in such manifolds \cite{Grana:2004bg} are but two examples.
\bigskip

{\bf Acknowledgement}: I am grateful to Martin Ro\v cek and Rikard von Unge checking the manuscript and 
helping with the teXing. The research is supported in part by VR grant 650-1998368.


\begin{thebibliography}{6666} 

\newcommand{\np}{{\em Nucl.\ Phys.\ }} 
\newcommand{\pr}{{\em Phys.\ Rev.\ }} 
\newcommand{\cmp}{{\em Commun.\ Math.\ Phys.\ }} 
\newcommand{\pl}{{\em Phys.\ Lett.\ }} 
%
\bibitem{Hitchin}
N.~Hitchin,
``Generalized Calabi-Yau manifolds,''
Q. J. Math.  {\bf 54}  (2003), no. 3, 281--308,
arXiv:math.DG/0209099.
%
\bibitem{Gualtieri}
M.~Gualtieri,
``Generalized complex geometry,''
Oxford University DPhil thesis, arXiv:math.DG/0401221.
%
\bibitem{Zumino:1979et}
B.~Zumino,
``Supersymmetry And Kahler Manifolds,''
Phys.\ Lett.\ B {\bf 87}, 203 (1979).
%
\bibitem{Alvarez-Gaume:1980vs}
L.~Alvarez-Gaume and D.~Z.~Freedman,
``Ricci Flat Kahler Manifolds And Supersymmetry,''
Phys.\ Lett.\ B {\bf 94}, 171 (1980).

\bibitem{Lindstrom:2004iw}
U.~Lindstrom, R.~Minasian, A.~Tomasiello and M.~Zabzine,
``Generalized complex manifolds and supersymmetry,''
arXiv:hep-th/0405085.
%
\bibitem{Grana:2004bg}
M.~Grana, R.~Minasian, M.~Petrini and A.~Tomasiello,
arXiv:hep-th/0406137.


\bibitem{Lindstrom:2004eh}
U.~Lindstrom,
``Generalized N = (2,2) supersymmetric non-linear sigma models,''
arXiv:hep-th/0401100.
%

\bibitem{courant}
T.~Courant,
``Dirac manifolds,''
{\it  Trans. Amer. Math. Soc.}  {\bf 319}  (1990), no. 2, 631--661. 
%

\bibitem{Gates:nk}
S.~J.~Gates, C.~M.~Hull and M.~Rocek,
``Twisted Multiplets And New Supersymmetric Nonlinear Sigma Models,''
Nucl.\ Phys.\ B {\bf 248} (1984) 157.
%

\bibitem{Schaller:1994es}
P.~Schaller and T.~Strobl,
``Poisson structure induced (topological) field theories,''
Mod.\ Phys.\ Lett.\ A {\bf 9} (1994) 3129
[arXiv:hep-th/9405110].
%

%
\bibitem{Lindstrom:2002jb}
U.~Lindstrom and M.~Zabzine,
``N = 2 boundary conditions for non-linear sigma models and Landau-Ginzburg
models,''
JHEP {\bf 0302} (2003) 006
[arXiv:hep-th/0209098].
%
\bibitem{Lindstrom:2002vp}
U.~Lindstrom and M.~Zabzine,
``D-branes in N = 2 WZW models,''
Phys.\ Lett.\ B {\bf 560} (2003) 108
[arXiv:hep-th/0212042].
%
\bibitem{Zabzine:2004dp}
M.~Zabzine,
``Geometry of D-branes for general N = (2,2) sigma models,''
arXiv:hep-th/0405240.

\bibitem{kapustin}
Anton Kapustin  and Yi Li
``Topological sigma-models with H-flux and twisted generalized complex  manifolds''
arXiv:hep-th/0407249 
 



\end{thebibliography}
\end{document}